\documentclass[5p,times]{elsarticle}
\usepackage{graphicx}
\usepackage{amsmath}
\usepackage[english]{babel}
\usepackage{amsfonts,amssymb}
\usepackage[mathscr]{eucal}

\usepackage{hyperref}
\hypersetup{colorlinks=true,linkcolor=blue,urlcolor=blue,citecolor=blue}

\begin{document}

\title{Mechanisms of FMR line broadening in CoFeB$-$LiNbO$_3$ granular films\\
in the vicinity of metal-insulator transition}

\author[ipp]{A.B.~Drovosekov\corref{cor}}
\ead{drovosekov@kapitza.ras.ru}
\author[ipp]{N.M.~Kreines}
\author[ipp,hse]{A.S.~Barkalova}
\author[nrc]{S.N.~Nikolaev}
\author[nrc,ire,itae]{V.V.~Rylkov}
\author[vstu]{A.V.~Sitnikov}

\cortext[cor]{Corresponding author}

\address[ipp]{P.L.Kapitza Institute for Physical Problems, RAS, 119334 Moscow, Russia}

\address[hse]{National Research University "Higher School of Economics", 101000 Moscow, Russia}

\address[nrc]{National Research Center "Kurchatov Institute", 123182 Moscow, Russia}

\address[ire]{Kotel'nikov Institute of Radio Engineering and Electronics RAS, 141190 Fryazino, Moscow region, Russia}

\address[itae]{Institute of Theoretical and Applied Electrodynamics RAS, 127412 Moscow, Russia}

\address[vstu]{Voronezh State Technical University, 394026 Voronezh, Russia}

\begin{abstract}
Metal-insulator (CoFeB)$_x$(LiNbO$_3$)$_{100-x}$ nanocomposite films with different content of the ferromagnetic (FM) phase $x$ are investigated by ferromagnetic resonance (FMR) technique. A strong change of the FMR line shape is observed in the vicinity of metal-insulator transition (MIT) of the film, where the hopping-type conductivity $\sigma$ modifies to the regime of a strong intergranular tunnelling, characterized by a logarithmic dependence $\sigma(T)$ at high temperatures. It is shown that below MIT, the FMR linewidth is mainly determined by the inhomogeneous distribution of the local anisotropy axes in the film plane. Above MIT, the contribution of this inhomogeneity to the line broadening decreases. At the same time, two-magnon magnetic relaxation processes begin to play a significant role in the formation of the linewidth. The observed behaviour indicates the critical role of interparticle exchange in the tunnelling regime above MIT of the nanocomposite.
\end{abstract}



\maketitle

\section{Introduction}

For many decades granular magnetic systems have attracted attention due to peculiar physical properties as well as various applications \cite{Dormann1992,Bedanta2013}. Magnetic nanoparticles find their use in a wide range of technological and interdisciplinary fields, including development of data storage devices \cite{Moser2002}, enhanced permanent magnets \cite{Sun2000}, magnetic resonance imaging \cite{Chung2004}, ferrofluids \cite{Torres2014}, biomedicine and health science \cite{Andra2007}.

The granular (CoFeB)$_x$(LiNbO$_3$)$_{100-x}$ metal-insulator nanocomposite (NC) is a synthetic multiferroic system that can be of great interest due to possibilities of non-trivial magneto-electric effects \cite{Udalov2018}. Recently, this NC was proposed as promising material for realizing resistive switching memory elements (memristors) for potential applications in neuromorfic networks \cite{Rylkov2018jetp,Nikiruy2019}. The ion beam sputtered films of such NC demonstrate an interesting structural special feature: the CoFeB metallic FM phase has a tendency to form essentially non-spherical granules inside the LiNbO insulator matrix. The FM particles have the shape of short nanowires elongated in the direction of the film growth (see Fig.~1(a) and Ref.~\cite{Rylkov2018}). The metal-insulator transition (MIT) in such NC was estimated as $x_c\approx43$~at.\%. Recently it was found that the perpendicular surface magnetic anisotropy of the CoFeB granules greatly affects the magnetic properties of this NC \cite{Rylkov2019}.

Studies of magnetization dynamics in metal-insulator NCs attract attention from both fundamental and applied points of view. The possibility to combine high resistivity and high magnetic permeability in granular systems makes them interesting for applications in high-frequency microelectronic devices, such as on-chip inductors and transformers \cite{Sullivan2009}. On the other hand, these structures are suitable objects to investigate the influence of individual particles parameters and interparticle interactions on dynamical characteristics of the system, in particular, magnetic relaxation mechanisms.

In this work, the (CoFeB)$_x$(LiNbO$_3$)$_{100-x}$ granular NC films are investigated by ferromagnetic resonance (FMR) technique. We study the behaviour of the FMR linewidth for films with different concentrations of FM phase below and above MIT $x_c\approx43$~at.\% to shed light upon the role of interparticle interactions in magnetic relaxation mechanisms.

Previously, the effects of percolation of the metal granules on the FMR characteristics were investigated in a number of works for different metal-insulator systems: Fe--SiO$_2$ \cite{Wang1995,Kakazei1999}, Ni--SiO$_2$ \cite{Kakazei1999}, Co--SiO$_2$ \cite{Gomez2004,Gomez2004B}, Co--Al$_2$O$_3$ \cite{Pires2006,Timofeev2012}, FeCo--Al$_2$O$_3$ \cite{Lesnik2003}. In contrast to these works, here we pay special attention to the region of concentration $x=44-48$~at.\%, \textsl{below} the percolation threshold of the investigated NC. In this region, the temperature dependence of the electrical conductivity was found to follow the unusual logarithmic law in a wide temperature range $T\approx10-200$~K \cite{Rylkov2018jetp}. It was recently shown theoretically \cite{Beloborodov2007,Efetov2003}, that such behaviour is not connected with the dimension of the system and the weak-localization-induced quantum correction, but can be explained by the renormalization of the Coulomb interaction, which affects the probability of electron tunnelling between granules. Formally, this case corresponds to the array of granules with a strong tunnel coupling between them \cite{Beloborodov2007,Efetov2003}. Below we present results of studies of FMR features in the magnetic granular systems in this interesting range of metal content near MIT.

\section{Samples and experimental details}

The NC (CoFeB)$_x$(LiNbO$_3$)$_{100-x}$ films with different content of FM phase $x=27-48$~at.\% were produced by the ion-beam sputtering onto glass-ceramic substrates using the composite targets of the parent metallic alloy Co$_{40}$Fe$_{40}$B$_{20}$ and oxide LiNbO$_3$ stripes placed onto the metal surface. The thickness of the sputtered films was about $d\sim760$~nm. The structural features of the samples were studied by transmission electron microscopy (TEM). The films consisted of the crystalline CoFeB granules, embedded into the amorphous matrix, with the in-plane size of 2--4~nm and elongated in the direction of NC growth up to 10--15~nm (see Fig.~1(a) and Refs.~\cite{Rylkov2018jetp,Rylkov2018} for details of samples preparation and structural characterization).

Ferromagnetic resonance was studied at room temperature on a laboratory-developed transmission-type spectrometer operating in the frequency range 7--36~GHz. The field-sweep FMR absorption spectra were measured in magnetic fields up to 17~kOe that could be applied at arbitrary angle with respect to the film plane (Fig.~1(b)). The ultra-high frequency magnetic field $h$ was oriented in the film plane perpendicular to the static field $H$.


\begin{figure}[t]
\centering
\includegraphics[width=.91\columnwidth]{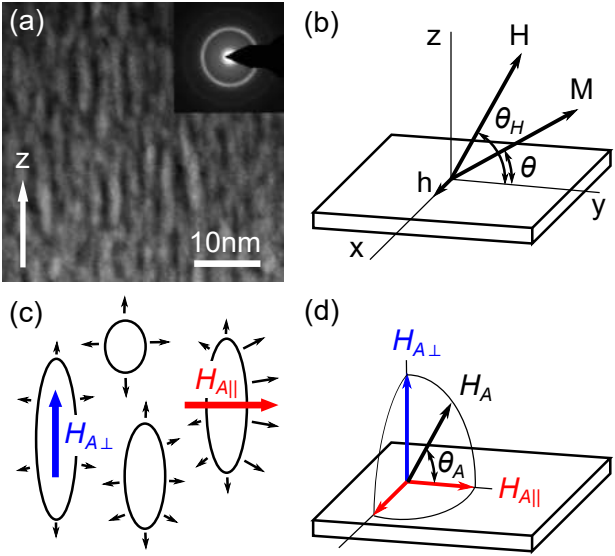}
\caption{a) Dark-field STEM image of the film cross-section ($x=48$~at.\%, from Ref.~\cite{Rylkov2018}). b) Experimental geometry. c)~Schematic representation of FM granules with random shapes. The surface anisotropy axes are shown by thin black arrows. The resulting effective anisotropy field of the granule is oriented along $z$-axis when the shape anisotropy is prevailing ($H_{A\perp}$), or in $xy$-plane when the inhomogeneous surface anisotropy is prevailing ($H_{A\parallel}$). d) A generalized case of ellipsoidal angular distribution of local easy axes in the film.}
\end{figure}

\begin{figure*}[t]
\centering
\includegraphics[width=\textwidth]{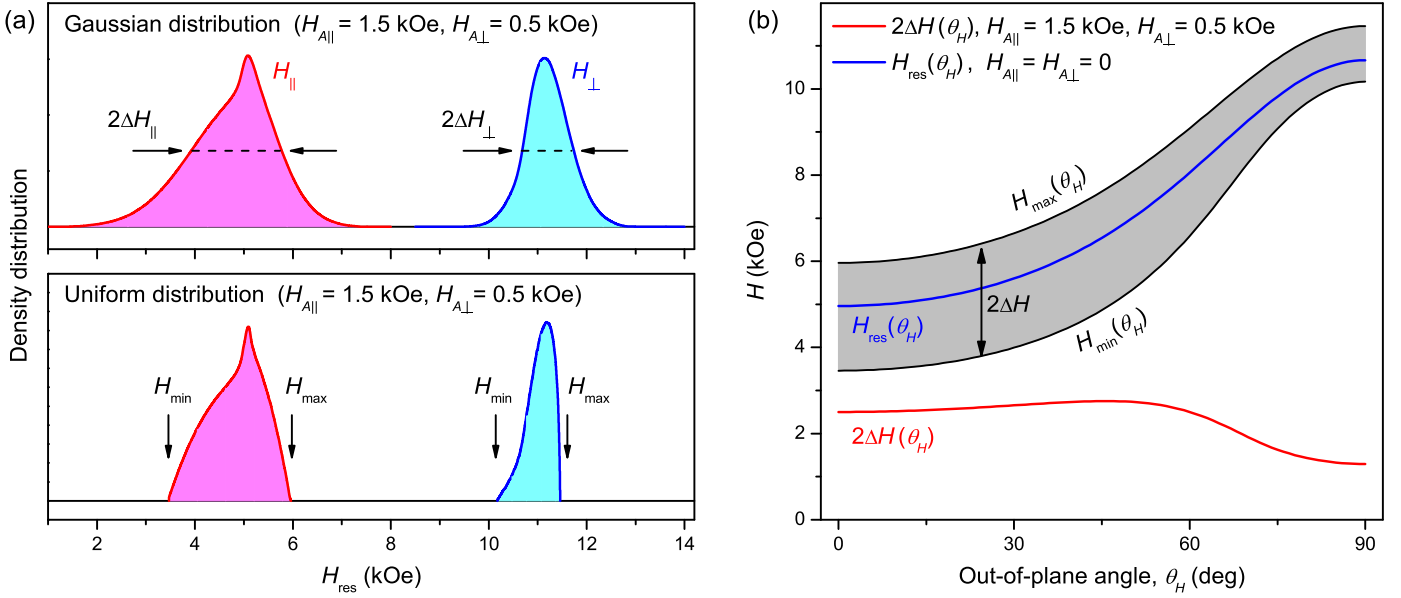}
\caption{a) Simulated distributions of local resonance fields in the granular film in parallel ($H_\parallel$) and normal ($H_\perp$) geometries. The upper and lower plots correspond to Gaussian and uniform distributions of the local easy axes respectively. The parameters are: $f=20.0$~GHz, $\gamma/2\pi=3.0$~GHz/kOe, $4\pi M_\mathrm{eff}=4.0$~kOe, $H_{A\parallel}=1.5$~kOe, $H_{A\perp}=0.5$~kOe. b) Calculated extremal $H_\mathrm{res}$ values as a function of $\theta_H$ for uniform distribution of the easy axes, Eq.~(10), and the resulting angular dependence $\Delta H(\theta_H)$. The curve $H_\mathrm{max}(\theta_H)$ is obtained using Eqs.~(13,14) in case $\mathbf{n}\parallel\mathbf{x}$. The curve $H_\mathrm{min}(\theta_H)$ is found with Eqs.~(13,15), numerically minimizing $H_\mathrm{res}$ in case $\mathbf{n}\perp\mathbf{x}$. The model parameters are the same as in the plot (a).}
\end{figure*}


\section{Theoretical background}

\subsection{FMR frequency in a thin film}

Ferromagnetic resonance condition for a thin homogeneous FM film is defined by well known equation (see for example \cite{Soohoo1965,Maksymowicz1992,Lindner2009})
\begin{equation}
\frac{\omega^{2}}{\gamma^{2}}=H_1 H_2,
\end{equation}
where $\omega=2\pi f$ is the frequency and $\gamma$ is the gyromagnetic ratio. When the magnetic field $H$ is applied at an angle $\theta_H$ with respect to the film plane (Fig.~1(b)), the effective fields $H_1$ and $H_2$ are given by
\begin{equation}
\begin{array}{l}
H_1 = H\cos(\theta_{H}-\theta)-4\pi M_\mathrm{eff}\sin^{2}\theta,\\
H_2 = H\cos(\theta_{H}-\theta)+4\pi M_\mathrm{eff}\cos2\theta,
\end{array}
\end{equation}
where angle $\theta$ defines the static orientation of magnetization vector with respect to the film plane and $4\pi M_\mathrm{eff}$ is the effective demagnetizing field including possible contribution of perpendicular uniaxial anisotropy.

The static orientation of the magnetization vector is defined by the equation
\begin{equation}
2H \sin (\theta_H -\theta) = 4\pi M_\mathrm{eff} \sin2\theta.
\end{equation}
Thus, solving Eqs. (1--3), the resonance field $H_\mathrm{res}$ can be found as a function of frequency $f$ and angle~$\theta_H$.

Note that according to \cite{Dubowik1996}, Eqs. (1--3) remain valid not only for homogeneous FM film, but also for the film formed by ideal periodic array of identical FM granules. In the real system, the random distribution of the granules shapes and inter-granular interactions may lead to inapplicability of Eqs. (1--3). However, if the fluctuations of the local effective fields are small, comparing with the net $4\pi M_\mathrm{eff}$ value, Eqs. (1--3) are still usable in the first approximation. At the same time, the presence of these fluctuations may have a significant effect on the line shape. We will see that such situation is realised experimentally in the considered granular films.

\subsection{FMR linewidth. Gilbert damping}

To describe the FMR linewidth we consider three mechanisms: intrinsic Gilbert damping, two-magnon scattering (TMS) and inhomogeneous line broadening \cite{Heinrich2005,Farle2013}. Approximately, their contributions to the total line half-width at half-maximum (HWHM), $\Delta H$, can be considered as additive:
\begin{equation}
\Delta H = \Delta H_\mathrm{G} + \Delta H_\mathrm{TMS} + \Delta H_\mathrm{inhom}.
\end{equation}
The Gilbert contribution is described by equation \cite{Lindner2009}
\begin{equation}
\Delta H_\mathrm{G} = \frac{\alpha\omega}{\gamma\Xi},
\end{equation}
where $\alpha$ is the Gilbert damping parameter \cite{Gilbert} and $\Xi$ is the so-called dragging function \cite{Lindner2009}. This function turns to 1 when the magnetization vector is oriented along the magnetic field ($\theta=\theta_H$), which is the case in parallel ($\theta_H=0^\circ$) and normal ($\theta_H=90^\circ$) geometries. For intermediate angles $0^\circ<\theta_H<90^\circ$ this function is \cite{Lindner2009}:
\begin{equation}
\Xi = \cos(\theta_H-\theta) - \frac{3H_1+H_2}{H_2(H_1+H_2)}H\sin^2(\theta_H-\theta),
\end{equation}
which leads to the increase of $\Delta H_\mathrm{G}$. According to Eq.~(5), the Gilbert contribution to the linewidth is equal in parallel and normal geometries and has a linear dependence on frequency.

\subsection{Two magnon mechanism of FMR line broadening}

To describe the TMS contribution, we consider the theory of surface inhomogeneities developed by Arias, Landeros and Mills \cite{Arias1999,Landeros2008} for ultrathin FM films. Within this theory, the TMS contribution to the linewidth is defined by expression
\begin{equation}
\Delta H_\mathrm{TMS} \approx \frac{\Gamma_\mathrm{TMS}}{\Xi}\arcsin\sqrt{\frac{H_1}{H_1 + 4\pi M_\mathrm{eff}}\frac{\cos2\theta}{\cos^2\theta}},
\end{equation}
where $\Gamma_\mathrm{TMS}$ is a cumbersome function of $H$ and $\theta_H$ and depends on the parameters of inhomogeneities (see Refs.~\cite{Lindner2009,Landeros2008}). Here, for simplicity, we will consider $\Gamma_\mathrm{TMS}$ as a constant, bearing in mind that its dependence on $H$ and $\theta_H$ is relatively weak.

The special feature of the considered TMS mechanism is that it contributes to the linewidth only for sufficiently small $\theta<45^\circ$. In particular, it means that $\Delta H_\mathrm{TMS}=0$ in normal geometry when $\theta=\theta_H=90^\circ$. In parallel geometry, Eq.~(7) gives approximately linear dependence of $\Delta H_\mathrm{TMS}$ on frequency \cite{Zakeri2007}.

Strictly speaking, Eq.~(7) is valid only for ultrathin ($d\lesssim10$~nm) films, while we deal with thick submicron FM films. Nevertheless, we will see that this relatively simple equation works sufficiently well in our case.

\subsection{Inhomogeneous FMR line broadening}

Fluctuations of shape and surface anisotropy of the granules may lead to inhomogeneous distribution of local anisotropy axes in the film (Fig.~1(c)). Considering the non-spherical form of the granules, we can expect that this distribution can also be non-spherical (Fig.~1(d)), so that the net effective anisotropy fields are different in cases when the local axis is oriented in the film plane and normal to the plane ($H_{A\parallel}$ and $H_{A\perp}$ respectively). To describe such ellipsoidal angular distribution of the effective anisotropy field $H_A$ we introduce the function:
\begin{equation}
\xi(H_A,\theta_A) = \left(\frac{H_A\cos\theta_A}{H_{A\parallel}}\right)^2 + \left(\frac{H_A\sin\theta_A}{H_{A\perp}}\right)^2,
\end{equation}
where $\theta_A$ is the angle between the easy axis and the film plane. Thus, the equation $\xi(H_A,\theta_A)=1$ describes a spheroid in the space of anisotropy vectors with semi-axes $H_{A\parallel}$ and $H_{A\perp}$. Then we may consider, for example, the Gaussian distribution of the easy axes with probability density
\begin{equation}
P_\mathrm{Gauss}(H_A,\theta_A) = P_0 \exp [-\xi(H_A,\theta_A)],
\end{equation}
where $P_0$ is the norming factor. The resulting distribution of the local resonance fields can be relatively easily found in the limit of high fields, when the magnetic moments of all the granules are oriented in the field direction. In this case, the local FMR frequency is defined by
\begin{equation*}
\begin{split}
& \frac{\omega^{2}}{\gamma^{2}} =  \left\lbrace H - H_M\sin^{2}\theta_H + H_A \left[ (n_y\cos\theta_H + n_z\sin\theta_H)^2 - n_x^2 \right] \right\rbrace \times \\
& \times \left\lbrace H + H_M\cos2\theta_H + H_A \left[ (n_y^2-n_z^2)\cos2\theta_H + 2n_yn_z\sin2\theta_H \right] \right\rbrace - \\
& - H_A^2 n_x^2 (n_y\sin\theta_H-n_z\cos\theta_H)^2,
\end{split}
\end{equation*}
where $H_M=4\pi M_\mathrm{eff}$ and $\mathbf{n}=(n_x,n_y,n_z)$ is the unit vector in the direction of the local easy axis. Examples of simulated distributions of $H_\mathrm{res}$ are shown in Fig.~2(a).

In general case of arbitrary fields, the procedure to simulate the shape of the line and to calculate the linewidth becomes too complicated. For this reason, we use a simplified approach. We consider the uniform distribution of the easy axes
\begin{equation}
P_\mathrm{uni}(H_A,\theta_A) =
\begin{cases}
P_0, & \mathrm{if} \quad \xi(H_A,\theta_A) \leq 1 \\
0, & \mathrm{if} \quad \xi(H_A,\theta_A) > 1,
\end{cases}
\end{equation}
and estimate the inhomogeneous broadening as
\begin{equation}
2\Delta H_\mathrm{inhom} \approx H_\mathrm{max}(H_A,\mathbf{n}) - H_\mathrm{min}(H_A,\mathbf{n}),
\end{equation}
where $H_\mathrm{max}(H_A,\mathbf{n})$ and $H_\mathrm{min}(H_A,\mathbf{n})$ are maximum and minimum resonance fields, depending on $H_A$ and $\mathbf{n}$. The proposed approach provides reasonable estimation of the inhomogeneous broadening (compare the upper and lower plots in Fig.~2(a)).

Note that if $H_{A\parallel}\rightarrow0$, then $H_A$ can be considered as a simple contribution to the effective easy-plane anisotropy $H_M=4\pi M_\mathrm{eff}$. In this situation, our approach is equivalent to the commonly used formula for inhomogeneous broadening \cite{Chappert1986}
\begin{equation}
\Delta H_\mathrm{inhom} = \left| \frac{\partial H_\mathrm{res}}{\partial H_M} \right| \Delta H_M \approx H_\mathrm{res}(H_M + \Delta H_M) - H_\mathrm{res}(H_M),
\end{equation}
where $\Delta H_M = H_{A\perp}$. In this case, the linewidth is maximal in normal geometry. In contrast, here we show that when $H_{A\parallel}$ exceeds $H_{A\perp}$, the linewidth in parallel geometry becomes larger than in normal geometry (see Fig.~2).

To estimate the inhomogeneous line broadening we need to find the extremal values of resonance field in Eq.~(11), taking into account the distribution (10). We note, that such values are achieved under the condition $\xi(H_A,\theta_A)=1$ when the easy axis is oriented either along the axis $x$ or in the plane $yz$ (see coordinate system in Fig.~1(b)). In these two cases, relatively simple expressions for FMR frequencies can be obtained in a standard way. In both cases, the formula for FMR frequency has the form similar to Eq.~(1):
\begin{equation}
\frac{\omega^{2}}{\gamma^{2}}=H_1^\prime H_2^\prime,
\end{equation}
where the effective fields $H_1^\prime$ and $H_2^\prime$ are modified as compared with $H_1$ and $H_2$ in Eq.~(2).

For the easy axis oriented along the axis $x$ (i.e. $\mathbf{n}\parallel\mathbf{x}$ and $H_A=H_{A\parallel}$), we consider the simplified case of relatively large fields, so that the magnetization vector lies in the plane $yz$. Then we obtain:
\begin{equation}
H_1^\prime = H_1-H_{A\parallel}, \qquad H_2^\prime = H_2.
\end{equation}
The condition of static equilibrium (3) remains true in this case.

In the second case, when the easy axis is oriented in the plane $yz$ (i.e. $\mathbf{n}\perp\mathbf{x}$), we obtain
\begin{equation}
\begin{array}{l}
H_1^\prime = H_1 + H_A \cos^2(\theta-\theta_A),\\
H_2^\prime = H_2 + H_A \cos2(\theta-\theta_A),
\end{array}
\end{equation}
where $H_A$ is a function of $\theta_A$ defined by equation $\xi(H_A,\theta_A)=1$. In this case, the static equilibrium condition (3) is modified to
\begin{equation}
2H \sin (\theta_H -\theta) = 4\pi M_\mathrm{eff} \sin2\theta + H_A \sin2(\theta-\theta_A).
\end{equation}
Using Eqs. (13--16), we can numerically find the extremal values of the resonance field, depending on the orientation of the anisotropy axis. As a result, with Eq.~(11), the inhomogeneous line broadening can be estimated as a function of frequency and orientation of the magnetic field. Fig.~2(b) illustrates the procedure to find the angular dependence $\Delta H_\mathrm{inhom}(\theta_H)$ for typical realistic parameters $H_{A\parallel}, H_{A\perp}$.


\begin{figure}[t]
\centering
\includegraphics[width=.82\columnwidth]{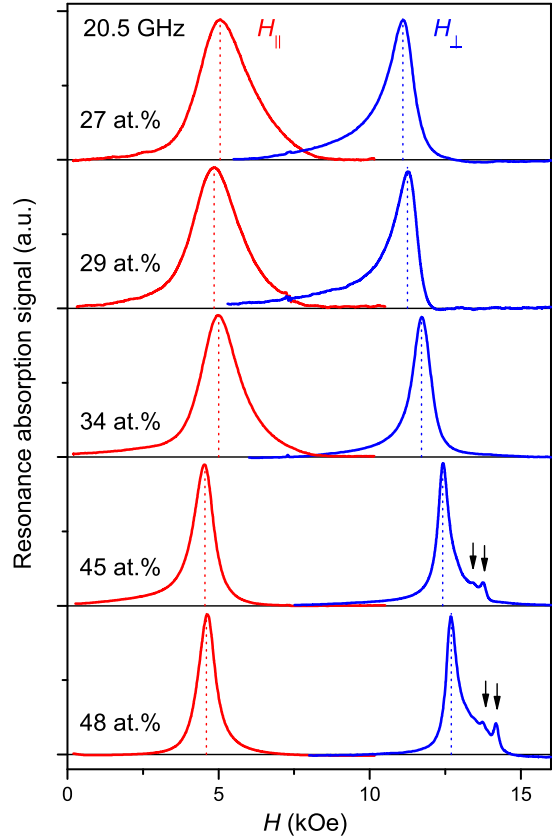}
\caption{Experimental spectra obtained in magnetic field applied parallel ($H_\parallel$) and perpendicular ($H_\perp$) to the film plane for samples with different content of the magnetic phase $x$. Arrows indicate additional weak peaks arising above MIT ($x_c\approx43$~at.\%). The frequency is 20.5~GHz.}
\end{figure}

\begin{figure*}[t]
\centering
\includegraphics[width=.97\textwidth]{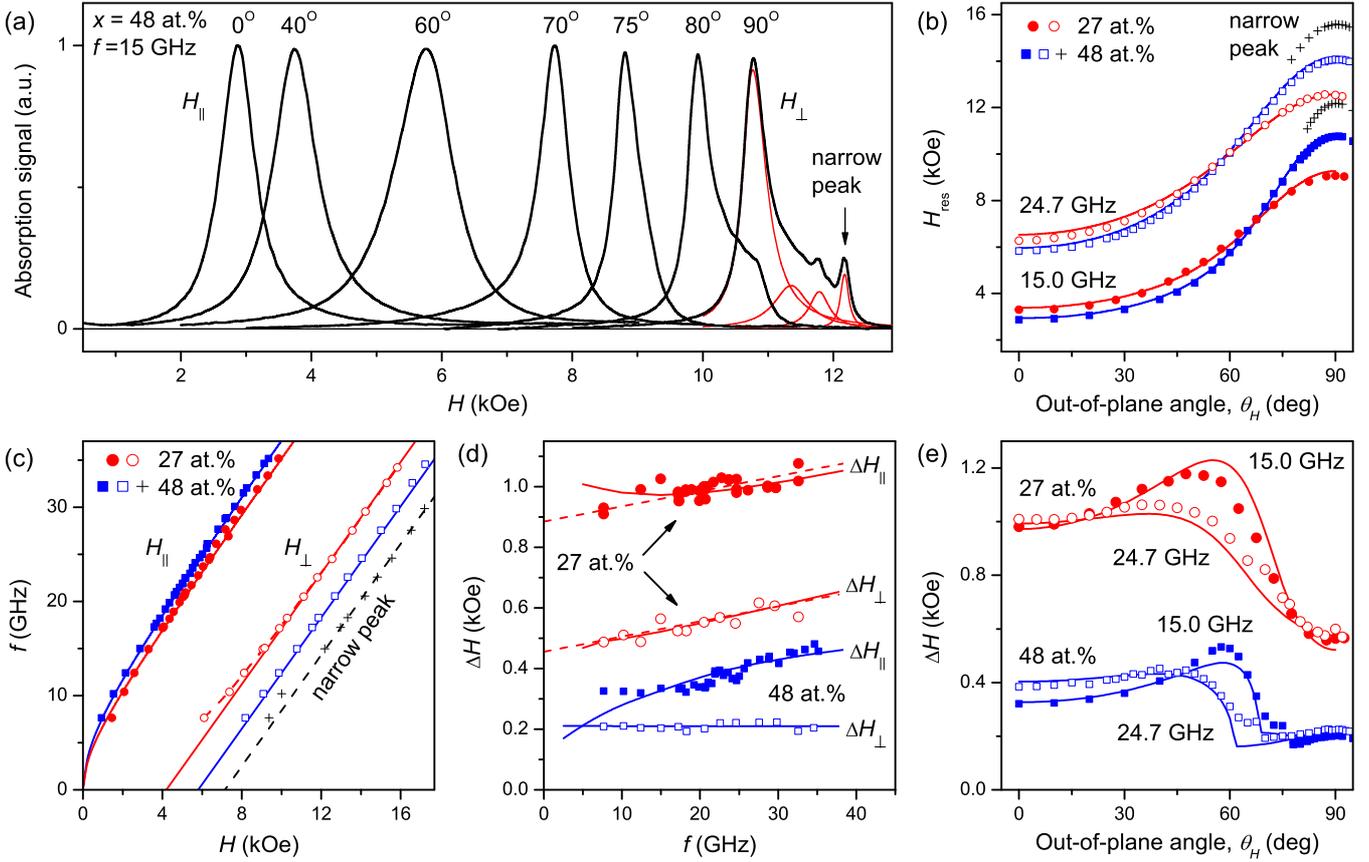}
\caption{a) Experimental spectra for NC with $x=48$~at.\% in magnetic field applied at different angles $\theta_H$ with respect to the film plane at frequency 15~GHz. b) Angular dependencies of the resonance field $H_\mathrm{res}(\theta_H)$ for films with $x=27$ and 48~at.\% at frequencies 15.0~GHz and 24.7~GHz. c) Resonance frequency \textsl{vs} field, $f(H)$, for films with $x=27$ and 48~at.\% in parallel and normal geometry. d) FMR HWHM linewidth $\Delta H$ as a function of frequency $f$ for films with $x=27$ and 48~at.\% in parallel and normal geometry. e) Angular dependencies of FMR linewidth, $\Delta H(\theta_H)$, for films with $x=27$ and 48~at.\% at frequencies 15.0~GHz and 24.7~GHz. Points in the plots (b--e) are experimental data, and the solid lines are theoretical approximation.}
\end{figure*}


\section{Results and discussion}

Fig.~3 shows resonance spectra for nanocomposites with different concentrations of the FM phase in two experimental geometries corresponding to parallel and perpendicular orientations of the magnetic field with respect to the film plane. We note that the position of the absorption maximum changes not very strongly for films in the investigated concentration range. The calculated effective demagnetizing field $4\pi M_\mathrm{eff}$ shows a weak monotonic increase from $4\pi M_\mathrm{eff}=4.2$~kOe for $x=27$~at.\% to $4\pi M_\mathrm{eff}=5.8$~kOe for $x=48$~at.\%. At the same time we observe a significant modification of the absorption line shape at increasing $x$. Samples with $x=27-34$~at.\%, i.~e. below MIT ($x_c\approx43$~at.\%), show broad lines with noticeable asymmetry which is a sign of magnetic inhomogeneity of the films \cite{Netzelmann1990}. For samples with $x=45-48$~at.\%, i.~e. above MIT, the absorption peak becomes much more narrow and symmetric, indicating a transition to homogeneous FM film. At the same time, for these samples, several additional weak narrow peaks appear in higher fields in perpendicular geometry.

Fig.~4(a) demonstrates FMR spectra for the sample with $x=48$~at.\% obtained at different angles $\theta_H$. It is seen that the additional peaks arise only in the vicinity of normal geometry. There are two possibilities to explain the origin of these peaks. First, they can be attributed to mesoscopic CoFeB clusters which begin to form in the films above MIT. The fact that these peaks disappear in the parallel geometry may be due to a significant broadening of these lines, so that they are not distinguished from the main peak. We note, that similar behavior was reported for granular Co--SiO$_2$ films in \cite{Pires2006}.

Another possible explanation of extra peaks in normal geometry is the excitation of surface resonance modes. Such modes may arise in the presence of surface anisotropy in case of quasi-homogeneous films with high content of FM phase. Similar behaviour was found, for example, for Co--SiO$_2$ \cite{Gomez2004,Gomez2004B} and Fe--SiO$_2$ \cite{Wang1995} granular films. The observation of spin-wave resonance modes was also reported in \cite{Kablov2016} for CoFeB--SiO$_2$ heterostructures. According to the theories by Soohoo \cite{Soohoo1963} and Puszkarski \cite{Puszkarski1979}, the surface modes can exist in normal geometry but disappear when the magnetic field deviates from the film normal. Such behaviour is possible in case of easy-plane-type surface anisotropy.

Figs.~4(b,c) demonstrate resulting angular and frequency dependencies of resonance field for samples with $x=27$ and 48~at.\%. For both samples the behaviour of the main peak is well described by theoretical formulas (1--3). A slight discrepancy between the experiment and the theory is observed only for NC with $x=27$~at.\% in the region of low frequencies in normal geometry (Fig.~4(c)). This discrepancy can be attributed to inhomogeneity of the film with low concentration of the FM phase. Indeed, in this case, the magnetic saturation is achieved only in relatively high fields, while in low fields the $4\pi M_\mathrm{eff}$ value can be reduced. At frequencies $f\gtrsim15$~GHz (i.e. in higher fields), the inhomogeneities seem to have negligible effect on the position of the FMR peak, while they strongly influence the line shape (see Fig.~3).

For the sample with $x=48$~at.\%, the behaviour of the main peak is well described by Eqs.~(1--3) in the entire frequency range. The experimental data for one most pronounced narrow additional peak in this film are also presented in Figs.~4(b,c).


\begin{table}[t]
\centering
\caption{Fitting parameters of the films.}
\vskip 3mm
\renewcommand{\arraystretch}{1.5}
\begin{tabular}{cccccc}
\hline
~ & $4\pi M_\mathrm{eff}$ & $H_{A\perp}$ & $H_{A\parallel}$ & $\alpha$ & $\Gamma_\mathrm{TMS}$\\
Film & (kOe) & (kOe) & (kOe) & ~ & (kOe)\\
\hline
$x=27$~at.\% & 4.2 & 0.32 & 1.03 & 0.018 & ---\\
$x=48$~at.\% & 5.8 & 0.36 & 0.12 & --- & 0.36\\
\hline
\end{tabular}
\end{table}


The resulting fitted $4\pi M_\mathrm{eff}$ values for two samples are shown in Tab.~1. The gyromagnetic ratio for both samples is $\gamma/2\pi\approx2.95$~GHz/kOe which corresponds to g-factor $g\approx2.1$ typical for CoFeB alloys \cite{Devolder2013,Glowinski2017}. It is interesting to note that $4\pi M_\mathrm{eff}$ value is not simply proportional to the concentration of the magnetic phase. Almost twofold decrease of concentration $x$ leads to only about 30\% decrease of the $4\pi M_\mathrm{eff}$ value. Thus, it seems that $M_\mathrm{eff}$ is not equal to the saturation magnetization of the film $M_S$. This fact can be explained by an additional anisotropic contribution that modifies $4\pi M_\mathrm{eff}$ as compared with the $4\pi M_S$ value:
\begin{equation}
4\pi M_\mathrm{eff}=4\pi M_S+H_U,
\end{equation}
where $H_U$ is effective anisotropy field. At first glance, there is an obvious reason for the additional anisotropic contribution due to the shape anisotropy of individual granules. According to the results by Dubowik \cite{Dubowik1996} in this case, $H_U$ would be
\begin{equation}
H_U=\Delta N^p M_p (1-x),
\end{equation}
where $M_p$ is the magnetization of the particle and $\Delta N^p=N_\perp^p-N_\parallel^p$ is the difference between its demagnetizing factors for the normal and in-plane directions. Taking into account that the granules are elongated in the direction of the film normal, we would expect a negative sign of $\Delta N^p\approx-2\pi$. Thus, the decreasing concentration $x$ should lead to more rapid decrease of $4\pi M_\mathrm{eff}$. This contradiction with the experiment suggests the presence of another easy-plane-type anisotropic contribution. The most probable origin of such anisotropy is the perpendicular surface anisotropy of the granules \cite{Rylkov2019} leading to preferable orientation of their magnetic moments in the film plane (this effect is illustrated schematically in Fig.~1(c)). This induced easy-plane anisotropy prevails at low concentrations $x$, while for higher $x$, it is suppressed by FM exchange between the granules \cite{Udalov2017} initiating magnetic homogenization of the film.

The role of granules anisotropies and interparticle interactions becomes more clear while comparing the behaviour of the FMR linewidth $\Delta H$ in films with $x=27$ and 48~at.\%. Fig.~4(d) shows the frequency dependencies of the linewidth for two samples in parallel and normal geometries (in case of $x=48$~at.\%, we analyse only the main absorption peak). The film with $x=27$~at.\% demonstrates approximately linear dependencies $\Delta H(f)$ in both geometries with about the same weak slope and large zero-frequency offset $\Delta H(0)$ (see dashed red lines in Fig.~4(d)). Such behaviour is typical for the case of inhomogeneous broadening with a weak contribution of the Gilbert damping mechanism \cite{Glowinski2017}. We note that in our case, the linewidth in parallel geometry is larger than in normal geometry. As we have shown above in theoretical section, this is a sign of significant in-plane distribution of the local easy axes ($H_{A\parallel}>H_{A\perp}$).

Another situation is observed for the film with $x=48$~at.\%. In this case, $\Delta H$ is constant in normal geometry but demonstrates increasing frequency dependence in parallel geometry. Such behaviour of the linewidth is typical for the TMS damping mechanism \cite{liu2011,Okada2017}.

Fig.~4(e) shows the angular dependencies of the linewidth for two samples. Here again, we clearly see qualitatively different behaviour of the curves $\Delta H(\theta_H)$ for films with different concentration $x$. The sample with $x=48$~at.\% demonstrates well defined plateau on the $\Delta H(\theta_H)$ dependence in the vicinity of $\theta_H=90^\circ$, which is a signature of the TMS mechanism of the FMR line broadening. The absence of such plateau for the sample with $x=27$~at.\% suggests that other mechanisms are prevailing.

Solid lines in Figs.~4(d,e) show the approximation of the experimental data, using the formalism described above in theoretical section. The fitting parameters are summarized in Tab.~1. As it may be seen, we achieved a good correspondence between the experiment and the model. In agreement with our qualitative analysis, the behaviour of the linewidth for the film with low $x=27$~at.\% is well described, considering inhomogeneous distribution of local anisotropy axes. We note (see Tab.~1) that the in-plane anisotropy fluctuations $H_{A\parallel}$ are about 3 times larger than the fluctuations of the normal component $H_{A\perp}$. In other words, the local anisotropy axes have a tendency to orient in the film plane. This result is difficult to explain, considering only the fluctuations of the granules shapes, because in this case we would expect $H_{A\perp} \gg H_{A\parallel}$. Thus, we suppose that the in-plane distribution of the anisotropy axes may arise from fluctuations of the surface anisotropy of the granules (see Fig.~1(c)).

For the sample with $x=48$~at.\% the $H_{A\parallel}$ value is strongly reduced but, instead, the TMS contribution arises. These effects can be explained by increasing FM exchange between the granules and their coalescence leading to suppression of individual particles anisotropies and formation of quasi-homogeneous FM film. Such exchange-induced crossover to the collective behaviour in an ensemble of anisotropic granules was investigated theoretically in a number of works (see, for example, \cite{Chudnovsky1986,Berzin2018}). There is still a noticeable fluctuating component of the perpendicular anisotropy $H_{A\perp}$ in the sample with $x=48$~at.\% (see Tab.~1). It can be attributed to some spread of demagnetizing fields $4\pi M_\mathrm{eff}$ in the film, for example, due to large scale fluctuations of the FM phase content $x$.

It is interesting to note the difference of the Gilbert damping for the investigated films. For the sample with $x=27$~at.\%, we found the Gilbert parameter $\alpha\approx0.018$. This value is close to the results of other authors for ultrathin (nanoscale) Co$_{40}$Fe$_{40}$B$_{20}$ films \cite{Devolder2013,Glowinski2017}. For the film with $x=48$~at.\%, the linewidth in normal geometry is frequency independent within the experimental accuracy. It means that the Gilbert parameter is much smaller for this sample, and can not be determined from our experiment. We note that the low values of Gilbert damping $\approx0.004$ were reported for bulk Co$_{40}$Fe$_{40}$B$_{20}$ films \cite{liu2011}.

\section{Conclusion}

We have studied the evolution of the position and shape of the FMR absorption peak with increasing content of the FM phase in (CoFeB)$_x$(LiNbO$_3$)$_{100-x}$ granular metal-insulator films. It is shown that in the investigated range of concentrations $x=27-48$~at.\%, the inhomogeneity driven mechanisms of the FMR line broadening are prevailing, while the intrinsic Gilbert damping contribution is small or negligible. In the limit of low $x$, the FMR spectrum can be described in terms of resonance of independent FM granules with random distribution of easy-axis anisotropies. We note that the perpendicular surface anisotropy of the granules plays an important role in formation of this distribution. At high $x$, the FMR linewidth is mainly determined by two magnon scattering processes.

The crossover between different relaxation regimes occurs at concentrations around $x\approx40$~at.\% which is close to the metal-insulator transition in the investigated nanocomposite ($\approx43$~at.\%). This crossover is clearly due to a critical change of interparticle exchange interactions. For $x\ll x_c$, these interactions are negligible as compared with granules anisotropies, while at $x>x_c$ an opposite situation takes place. In the latter case, a strong tunnel coupling between the granules initiates a formation of quasi-homogeneous ferromagnetic film. We note that the existence of such coupling in films with $x=44-48$~at.\% is confirmed by the logarithmic temperature dependence of their conductivity \cite{Rylkov2018jetp}.

\section*{Acknowledgments}

This work was partially supported by the Russian Foundation for Basic Research (projects 18-07-00772, 18-07-00756, 19-07-00471), and by the Basic Research Program of the Presidium of Russian Academy of Sciences 1.4. "Actual problems of low temperature physics".

The research work on the synthesis and structural characterization of NC (CoFeB)$_x$(LiNbO$_3$)$_{100-x}$ films was supported by the Russian Science Foundation (Grant No. 16-19-10233).

\bibliography{Manuscript}

\end{document}